# Strangeness Report

Federico Antinori

Istituto Nazionale di Fisica Nucleare, Padova, Italy

**Abstract**

The paper provides a short report on strangeness production in ultrarelativistic nucleus-nucleus collision, with the main stress on strange particle abundances.

## 1. Introduction

Strangeness is a vast subject. The amount of measurements relative to strange particles is huge. A major reason why strangeness is so interesting is that the strange flavour is not present at the beginning of the collision; it has to be produced by "cooking" in the reaction. The study of strangeness production, therefore, provides us with essential information about the physical environment created in nucleus-nucleus collisions.

The primary piece of information is just *how much* strangeness is produced, i.e. the strangeness abundance. In these proceedings, many measurements involving strange particles are covered in other review papers [1-5]; I will concentrate mainly on the abundance of strange particles: particle yields and particle ratios.

Our present knowledge on the pattern of enhancement of the production yields of strange and multi-strange particles in nucleus-nucleus collisions is summarised in section 2. Section 3 contains a brief discussion of the status of the comparison between experimental data and conventional hadronic transport models. Thermal equilibrium fits are discussed in section 4, while the experimental observations which may point to physics beyond equilibrium thermodynamics are discussed, rather speculatively, in section 5. The $\Phi$ mysteries are treated separately in section 6. Finally, I offer my conclusions in section 7. My oral presentation at the Quark Matter 2004 Conference also contained a selection of results on spectra from RHIC which could not find their way into the proceedings for reasons of space.

## 2. Strangeness enhancement pattern

In the Quark-Gluon Plasma (QGP) we expect deconfinement to be accompanied by a partial restoration of the chiral symmetry. As chiral symmetry is restored, the mass of the strange quark is expected to decrease from its constituent value to its current value of about 150 MeV, of the same order of the critical temperature. Therefore we expect abundant production of strange quark-antiquark pairs, mainly by gluon-gluon fusion [6,7]. As the partons are deconfined, we expect this enhancement of strangeness production to be more pronounced for multi-strange particles, which can now be built using uncorrelated strange quarks which were produced in independent microscopic interactions: we therefore expect the enhancement to increase with the particles' strangeness content [8]. Indeed, such a pattern of strangeness enhancement has been observed by the WA97 and NA57 Collaborations in Pb-Pb collisions at the SPS at a beam momentum of 158 A GeV/*c* ($\sqrt{s_{NN}}$ = 17.3 GeV) [9,10,11]. The enhancement is defined as the yield per wounded nucleon relative to the yield per wounded nucleon in a light reference system (for instance pp or pBe): for a generic collision system AB the enhancement of a generic particle j is computed as:

$$E_j = \frac{(Y_j / N_{wound})_{AB}}{(Y_j / N_{wound})_{ref}} \qquad (1)$$

where $Y_j$ is the yield of the particle in the phase space window under consideration and $N_{wound}$ is the estimated number of wounded nucleons (i.e. nucleons expected to undergo a primary inelastic collision). In the case of WA97 and NA57 the reference system is pBe. As can be seen from Figure 1, no enhancement is observed in pPb collisions, while in PbPb collisions one observes clear enhancements, which are found to increase with centrality and strangeness content, reaching about a factor 20 for the |s|=3 Ω baryons for the most central collisions.

As of this conference, we know that a qualitatively similar pattern is also observed at RHIC (by the STAR Collaboration in Au-Au collisions at $\sqrt{s_{NN}}$ = 200 GeV [12]) and at lower SPS energy (by the NA57 Collaboration in Pb-Pb collisions at a beam momentum of 40 A GeV/*c*, $\sqrt{s_{NN}}$ = 8.8 GeV [11]). I will come back to these later.

I would like to take a moment to discuss in more detail the issue of strangeness enhancement in pA.

Within a simple two-component model of particle production in pA collisions – i.e. no enhancement in pA – one expects particle yields to scale with the number ν of proton-nucleon inelastic collisions as:

$$Y_{pA} = (\nu/2 + 1/2) Y_{pp} = N_{wound} Y_{pp} / 2 \qquad (2)$$

(in pA $N_{wound}$ = ν + 1).
In case of enhancement:

$$(Y/N_{wound})_{pA} = E \cdot (Y/N_{wound})_{pp} \qquad (3)$$

or, equivalently,

$$Y_{pA} = E \cdot (\nu/2 + 1/2) Y_{pp} \qquad (4)$$

As discussed above, no such enhancement is observed from pBe to pPb collisions.

For completeness, I will discuss an alternative definition of the enhancement in proton-nucleus collisions [13]. There, the enhancement is "blamed" only on the projectile component: a new definition of enhancement is given (I will denote it by F) such that:

$$Y_{pA} = (\nu/2 + F/2)Y_{pp} \qquad (5)$$

(this of course implies that strange quarks can somehow be divided into *projectile* and *target* strange quarks).
The relation between E and F is then given by:

$$(F-1) = (E-1)N_{wound} \qquad (6)$$

As can be seen, with the new definition an enhancement within pA appears even if the pA yields are proportional to $N_{wound}$, as is found to be the case experimentally (Figure 1). I shall stick with the usual definition of enhancement (eq. 4).

## 3. Hadronic transport

Hadronic transport codes do reasonably well on singly strange particles, but fail to reproduce the production of multi-strange particles at SPS and RHIC. A number of studies have been carried out on the subject (see, for instance, [14-17]). There's an interesting recent development on this issue: Huovinen and Kapusta have solved a network of rate equations for baryon abundances in a hadronic system [17]. They start from the critical temperature $T_c$ and evolve hyperon abundances as the system expands hydrodynamically and the temperature drops. If they start from a zero amount of hyperons at $T_c$, they find that the abundances stay below the thermodynamical equilibrium values until the system has cooled down to 120 MeV or so: hadronic reactions are too slow to reach equilibrium, whereas, as we shall see in a minute, experimentally particle abundances are found to be close to the thermodynamical equilibrium values corresponding to a temperature as high as 170 MeV. In case the calculation is repeated starting from the thermodynamic equilibrium abundances at $T_c$, it is found that during the subsequent evolution, as the system expands and cools, the hyperon abundances always remain above the equilibrium values: hadronic rates cannot even keep up with the expansion rate.

Hadronic transport does better if the masses are reduced towards the chiral values – or, somewhat equivalently, the string tension is enhanced with respect to the standard $e^+e^-$/pp/pA value – and/or the contribution from inelastic scattering during the expansion is enhanced beyond the known behaviour of hadronic cross sections. We haven't totally given up the hope of reproducing the hyperon enhancements within a purely hadronic realm as yet (see, for instance, [18,19]), but hadronic models are certainly getting more and more exotic.

Meanwhile, there is a nice new measurement coming from the AGS: the E895 Collaboration has measured the production of the $\Xi^-$ close to threshold in AuAu collisions at a beam momentum of 6 A GeV/*c*. The measured multiplicity of $\Xi^-$ and its centrality dependence can be reproduced with the RQMD code [20]: it seems therefore that hadronic transport still works fine at the AGS, but not any more at SPS and RHIC energies.

## 4. Thermal fits

Thermal fits on the contrary don't do badly at all. The latest in the series, the brand new RHIC 200 GeV fit including multi-strange particles, was presented at this meeting [21]. Once more, such a fit is in reasonably good agreement with the data: the relative particle abundances are found to be close to those expected at thermodynamical equilibrium for a grand-canonical system, even for the rare multi-strange particles. It is interesting to look at the energy dependence of the chemical freeze-out temperature extracted from such fits (Figure 2). Back in the 60s it had been realised [23, 24] that if the resonance mass spectrum grows exponentially (as seems indeed to be the case), there is a maximum possible temperature for a system of hadrons. And indeed (see Figure 2) we do not seem to be able to observe a system of hadrons with a temperature beyond a maximum value of the order of 170 MeV. (From the RHIC 200 fit mentioned above [21], the STAR Collaboration extracts a chemical freeze-out temperature $T_{ch} = 160 \pm 10$ MeV, this point is not included in Figure 2).

The extracted freeze-out points at SPS and RHIC lay on the QCD phase diagram very close to the predicted QGP phase boundary (see for instance [25]). This is of course just what one would expect if the system's chemical composition were essentially established at QGP hadronization, with little subsequent modification during the hadronic phase.

There is an important distinction to be made between canonical and grand-canonical equilibrium. It has to do with the energy penalty that has to be paid in order to introduce a strange particle into the system. In the canonical treatment it has to be computed taking into account also the energy which must be spent to create a companion in order to ensure strangeness conservation, while in the grand-canonical limit one needs only consider the energy necessary for the creation of the particle itself: the rest of the system acts as a reservoir and effectively "picks up the slack". The estimated effect of the removal of canonical suppression increases with strangeness and is of the order of the observed PbPb vs pBe hyperon enhancements [26]. The detailed centrality dependence of the enhancements, however, is not reproduced (although it must be said that the modelling of the collision centrality in [26] is very crude).

Does this provide us with an *explanation* for the observed enhancement pattern? If a system in thermodynamical equilibrium is very large, it will be in grand-canonical equilibrium, but the reverse is not necessarily true: having a large system, per se, is not a sufficient condition for having grand-canonical equilibrium. If, for instance, nucleus nucleus collisions were just a simple superposition of pp collisions, they would have to be treated canonically all the same, in spite of the system being large. In a certain sense in order to be grand-canonical a system must also *know* it is large: it must somehow know that an $\overline{\Omega}^+$ generated at one point can be compensated by, say, an $\Omega^-$ on the other side of the fireball. To be more precise: what counts is the size of the correlation volume: the volume over which strangeness is exactly conserved. "Canonical suppression is removed", then, is not an explanation, but just an observation; a synthetic way of describing what we see.

So, the question remains: how is canonical suppression removed? How can information travel so quickly through the system? Conventional hadronic transport, as discussed above, is too slow. Such fast equilibration is on the other hand just what one would expect in case of deconfinement. I would like mention two recent ideas on the subject. One, quantum coherence in the fireball, was put forward by Stock [27]: the long-range correlations needed for grand-canonical equilibrium to appear may be due to the non-locality of Quantum-Mechanical superposition, as the impinging nucleons undergo rapid sequential collisions with their target counterparts. I don't know whether this is being developed into a testable model; it would be interesting to know what one should expect for pA collisions. The other is illustrated in a paper by Braun-Munzinger, Stachel and Wetterich [28]: in a system of hadrons close to $T_c$ the density is expected to become very large; collective behaviour may then appear in the form of multi-particle collisions. These are expected to be important only very close to the phase boundary; as an example, in this model the equilibration time of the $\Omega^-$, close to $T_c$, scales approximately as $T^{-60}$. If this is true, thermal fits actually provide a way of experimentally measuring $T_c$ (this is indeed one of the main conclusions in [28]). The model in fact provides a suggestion for a microscopical mechanism to restore particle abundances to the hadronic equilibrium values as the system re-hadronizes.

## 5. Beyond equilibrium?

In order to look for signs of non-equilibrium, one must turn to fine details. Figure 3 shows the result of a thermal fit to NA49 158 A GeV/c data [29]. The agreement is very good, except for the $\Lambda(1520)$, which deviates from the systematics roughly by a factor two. An explanation of this disagreement has been proposed based on the existence of medium effects on the $\Lambda(1520)$ [30] (this actually also goes to show the degree of confidence in thermal fits we now have…). An analogous situation is seen at RHIC, where again resonances display the largest deviations from the thermal fit systematics [21]. These deviations are again attributed to medium effects and are exploited to study the relative importance of resonance rescattering and regeneration [1].

Sometimes (such as in the case of the NA49 fit just discussed [29]) it is preferred to use $4\pi$ particle ratios as input for the thermal fits, in order to be safe in case strangeness ends up too far away in rapidity from where it was originally produced (of course one has then to assume a global equilibrium with same temperature and, especially, same baryon density at different rapidities). In this case, it is necessary to introduce a strangeness undersaturation parameter $\gamma_s$, such that the thermal equilibrium abundance of each particle is modified by a factor $\gamma_s^{N(s+\bar{s})}$ where $N(s+\bar{s})$ is the number of strange quarks plus strange antiquarks contained in the particle. Experimentally, the $4\pi$ fits of NA49 for Pb-Pb collisions at the SPS require a strangeness undersaturation parameter $\gamma_s \sim 0.7 \div 0.8$. The Rafelski model (see for instance [31]) also contains a second non-equilibrium parameter – $\gamma_q$ – which controls the overall abundance of $q\bar{q}$ pairs. With respect to the previous situation, $\gamma_s$ becomes $\gamma_s/\gamma_q$, and an additional factor $\gamma_q^B$ modifies the abundance of baryons. The fits get very good (see [31] for RHIC 130 and [29] for the SPS). At RHIC 130 one finds [31] $\gamma_q = 1.6 \pm 0.2$ and $\gamma_s/\gamma_q = 1.3 \pm 0.1$: it looks like strangeness may actually be enhanced above the hadronic equilibrium values. It is worth noting that pentaquark production (if thermal!) would be particularly sensitive to the need for introducing $\gamma_q$: the abundance of pentaquarks being controlled by an additional $\gamma_q^2$ factor [29,32].

Figure 4 shows a comparison of the $\Xi^-$ enhancements at 40 and 160 A GeV/c from NA57 [11]. For the most central bin, the enhancement is larger at 40 GeV than at 160 GeV. Qualitatively, an increase of the enhancement at low energy is actually predicted by the canonical suppression model, the magnitude of the observed effect, however, is very small compared with the predicted increase by a factor 4 to 5 [33]. The centrality dependence of the enhancement at low energy is steeper than at the top SPS energy. Putting the two observations together, the possibility comes to mind that at 40 GeV the yield may be rising fast with centrality towards the grand-canonical value, but does not go all the way.

The energy dependence of the K/π ratios in nucleus-nucleus collisions is plotted in Figure 5 [34]. The $K^+/\pi^+$ ratio (Figure 5, left) has a change of behaviour around a beam momentum of 30 A GeV/c ($\sqrt{s_{NN}} = 7.62$ GeV). It has been suggested that such behaviour could signal the onset of the chiral transition (for the latest, see [30]). The reasoning goes as follows: as long as one is dealing with the constituent value of the strange quark mass, increasing the collision energy increases the $K^+/\pi^+$ probability. As the quark masses are restored to their chiral value, the $K^+/\pi^+$ ratio should instead become independent of the collision energy. The $K^-/\pi^-$ ratio (Figure 5, right), however, has a smooth, monotonic behaviour with energy. When interpreting these data, one has to take into account that, as the collision energy is increased, one creates systems with lower and lower baryon density; the $K^+(\bar{s}u)$ and $K^-(s\bar{u})$ abundances are sensitive to the baryon density, as can be seen from Figure 6, where the $K^-/K^+$ ratio is plotted as a function of the antiproton/proton ratio for a variety of measurements [35]. Data from the BRAHMS Collaboration shown at this conference [36] actually indicate that at RHIC as the rapidity (and therefore the baryon density) is increased, the $K^+/\pi^+$ ratio goes up and the $K^-/\pi^-$ ratio goes down, a similar behaviour as that obtained by decreasing the energy from the RHIC to the SPS range. It has actually been shown that, based on the thermal model and on the experimental chemical freeze-out curve, one would actually expect the Wroblewski factor $\lambda_s$ (which measures the relative abundance of strange versus light newly-produced quark-antiquark pairs), to have a maximum around a beam momentum of 30 A GeV/c [37]. On the basis of such calculations, one would indeed expect a broad maximum in the $K^+/\pi^+$ ratio around low SPS energy, as shown by the dashed line in Figure 5, but nothing as sharp as the observed "horn", which is, therefore, as of yet not really understood.

I would like to conclude this section by discussing an interesting regularity (see Figure 7). While the absolute values of the midrapidity yields of $\bar{\Lambda}$ and $\bar{\Xi}^+$ for central heavy-ion collisions increase strongly with energy, the absolute values of the $\Lambda$ and $\Xi^-$ yields remain pretty constant as $\sqrt{s_{NN}}$ ranges from 8.8 to 130 GeV [38]. The statistical fit parameters – the temperature T and the baryochemical potential $\mu_B$ (and – when used – also the non-equilibrium parameters $\gamma_s$ and $\gamma_q$) – are only sensitive to particle ratios. The overall normalizations are controlled by a separate volume parameter V. It looks as though T, $\mu_B$ and V all vary with energy, but in such a way as to ensure that the $\Lambda$ and $\Xi^-$ yields stay approximately constant. There must be a simple reason for this, but within thermal models this seems to be just an accident.

## 6. The mysterious Φ

In order to disentangle possible effects of strangeness undersaturation and canonical suppression the Φ could provide a crucial tool. Since the Φ is not openly strange, it should not feel the effects of canonical suppression; on the other hand in case of strangeness undersaturation its abundance is expected to be suppressed by a factor $\gamma_s^2$. Unfortunately, however, our knowledge of Φ production is blurred by an apparent discrepancy between NA49 (which detects the Φ via the $K^+K^-$ decay down to almost zero transverse momentum) and NA50 (which detects the Φ via the $\mu^+\mu^-$ decay, at transverse momenta above 500 MeV or so) concerning both the yield and transverse mass slope of the Φ [39,40].

The NA60 experiment, which should provide a high statistics sample of low transverse momentum $\Phi \to \mu^+\mu^-$) is expected to shed some light on this issues (NA60 is actually also looking at the $K^+K^-$ channel). A preliminary $\Phi \to \mu^+\mu^-$ signal from a small subsample of peripheral events collected during the recent InIn run has been shown already at this conference [41]. Further information should come from the PHENIX Collaboration, which has already extracted a $\Phi \to K^+K^-$ signal and is expected to soon provide a $\Phi \to e^+e^-$ signal as well [42].

Finally, NA49 reports a small (0.5 - 1 MeV) deviation from the Particle Data Group tables' mass [43] in the three PbPb data sets for which they have extracted a Φ signal, but not in pp [44]. This observation has not yet been independently verified.

## 7. Conclusions

The study of the production of strangeness in ultrarelativistic nucleus-nucleus collisions provides evidence for a high degree of chemical equilibration, even for the rare multistrange particles. The freeze-out curve extracted from thermal equilibrium fits merges with the expected QGP phase boundary at SPS and RHIC. A rapid equilibration of strangeness is indeed one of the originally predicted QGP signatures, while hadronic transport does not seem to be able to provide for such a degree of equilibration within the short collision time.

The chemical temperature at freeze-out appears to saturate at a value of 170 MeV, which is reached in the SPS energy range.

RHIC is opening up new windows for strangeness, with results on $R_{AA}$ and elliptic flow for multi-strange particles, strange and multistrange particles at high $p_T$, pp reference data with very high statistics. We should be ready for surprises.

There are indications that strangeness may actually be enhanced above hadronic equilibrium. This has to be clarified: we need the highest possible precision data from RHIC on particle yields and particle ratios (and possibly theory to provide an agreed unique statistical approach). We obviously need to understand the Φ mysteries.

Finally, I think we need to understand what is going on in the region $\sqrt{s_{NN}} = 5 - 10$ GeV, where the freeze-out curve meets the phase boundary, and there is indication of anomalous behaviour. This could be done with further low energy running at the SPS or possibly with the SIS 200 machine at GSI.

## Acknowledgements

I would like to thank Francesco Becattini, Rene Bellwied, Giuseppe Bruno, Helen Caines, Domenico Elia, Marek Gazdzicki, Adam Jacholkowski, Carlos Lourenço, Djamel Ouerdane, Emanuele Quercigh, Jan Rafelski, Karel Šafařík, Reinhard Stock and Julia Velkovska who helped me in the preparation of this report with information, discussions and criticism.

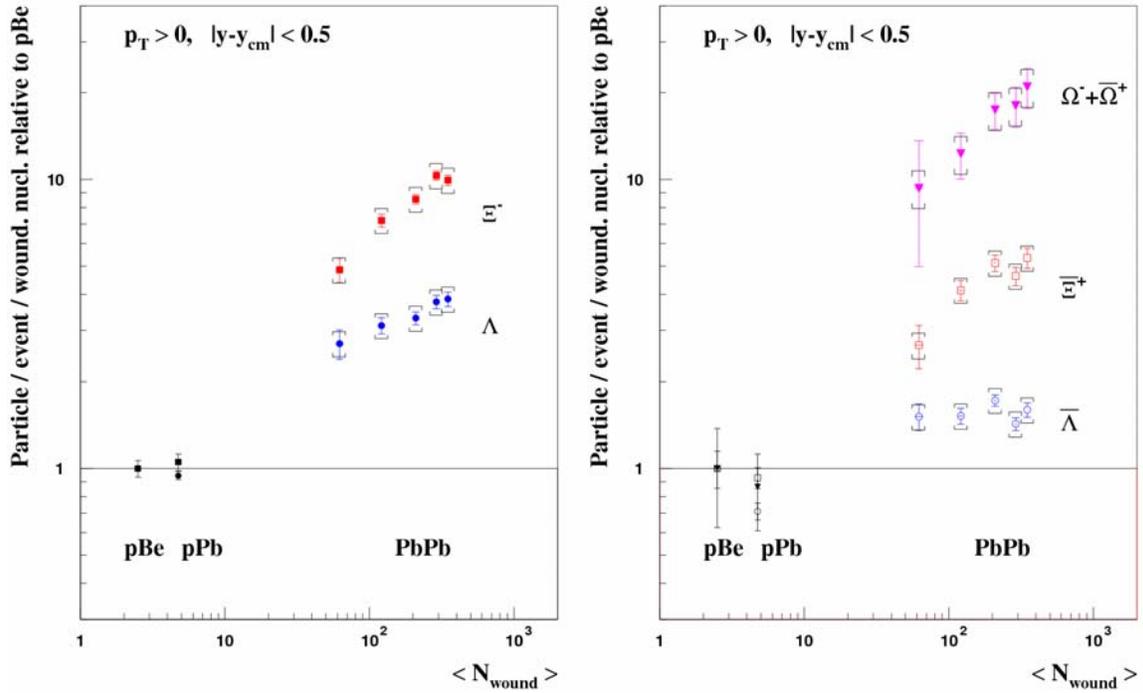

Figure 1: Pattern of enhancements of strange and multi-strange baryons at the top SPS energy (NA57).

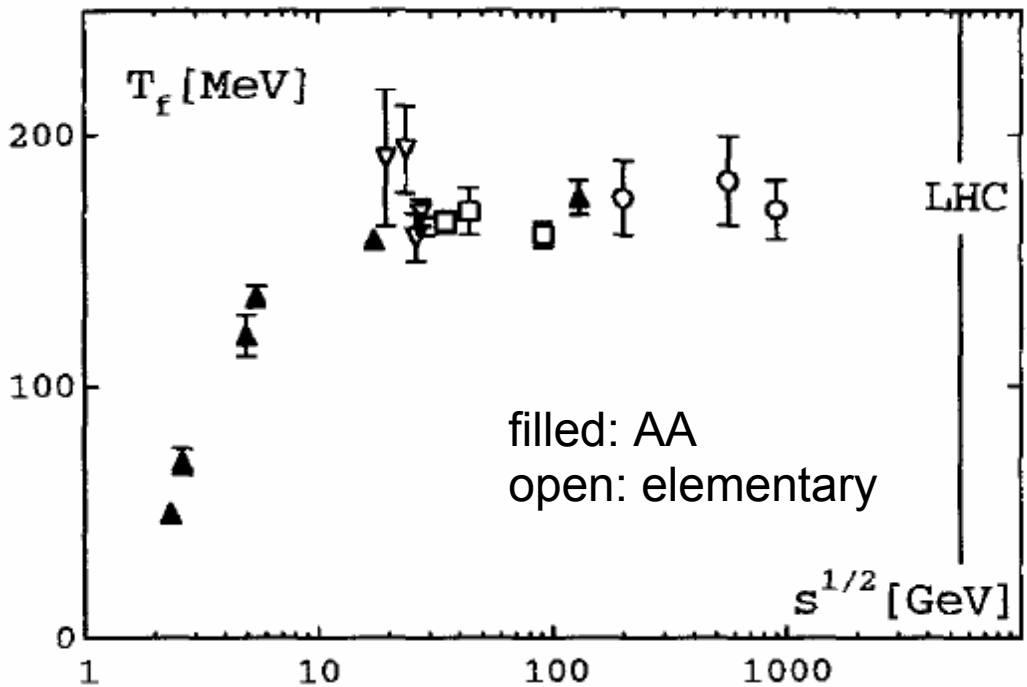

Figure 2: Hadronization temperature vs c.m.s. energy for e+e- (open squares), pp (open triangles), pp (open circles) and AA collisions (filled triangles) [22].

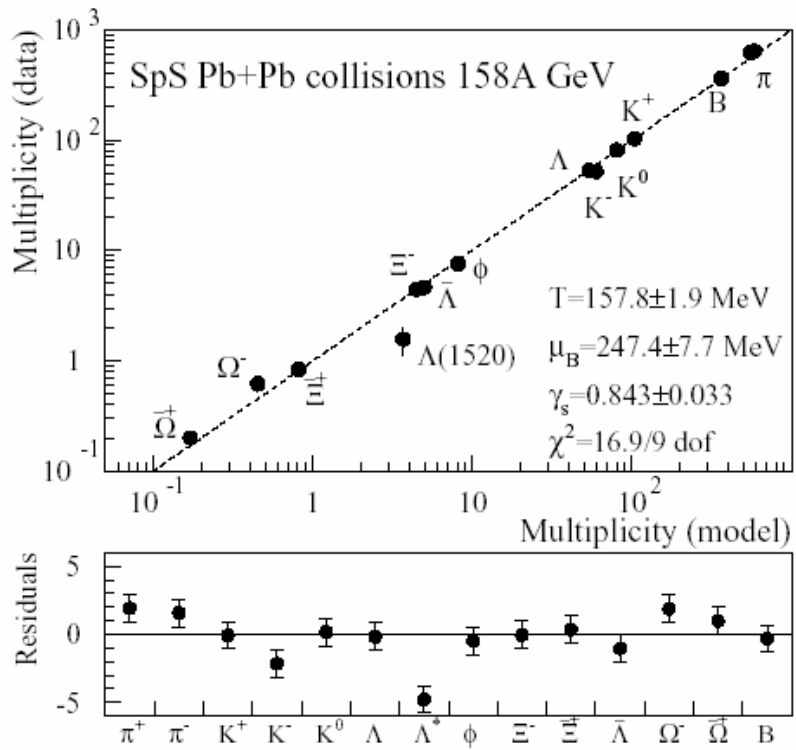

Figure 3. Thermodynamic equilibrium fit to NA49 data [29]

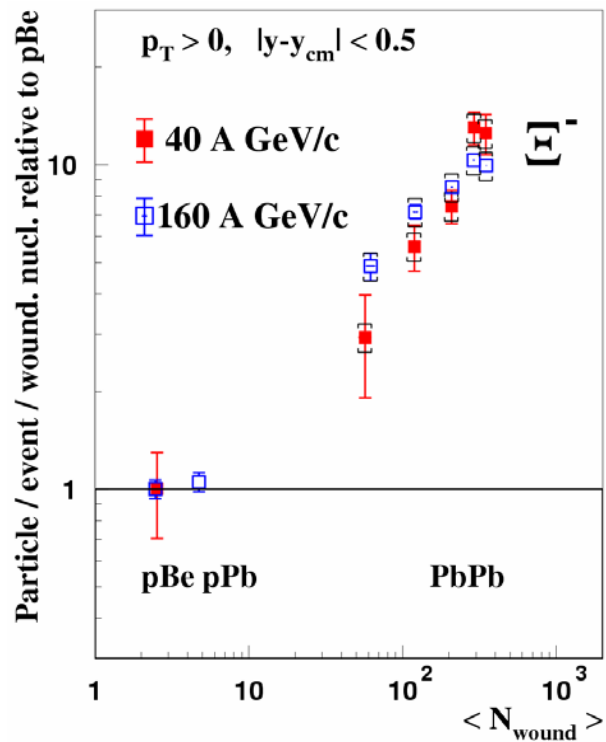

Figure 4. Comparison of the $\Xi^-$ enhancements at 40 and 160 A GeV/$c$ (NA57) [11]

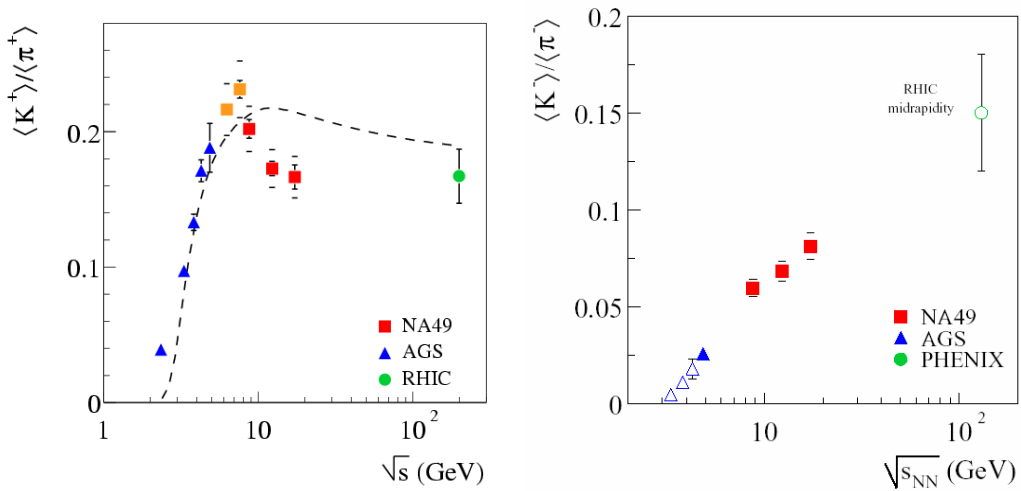

Figure 5. Energy dependence of the $K^+/\pi^+$ (left) and $K^-/p^-$ (right) ratios in nucleus-nucleus collisions. The dashed line on the right-hand plot indicate the thermal model expectation [34].

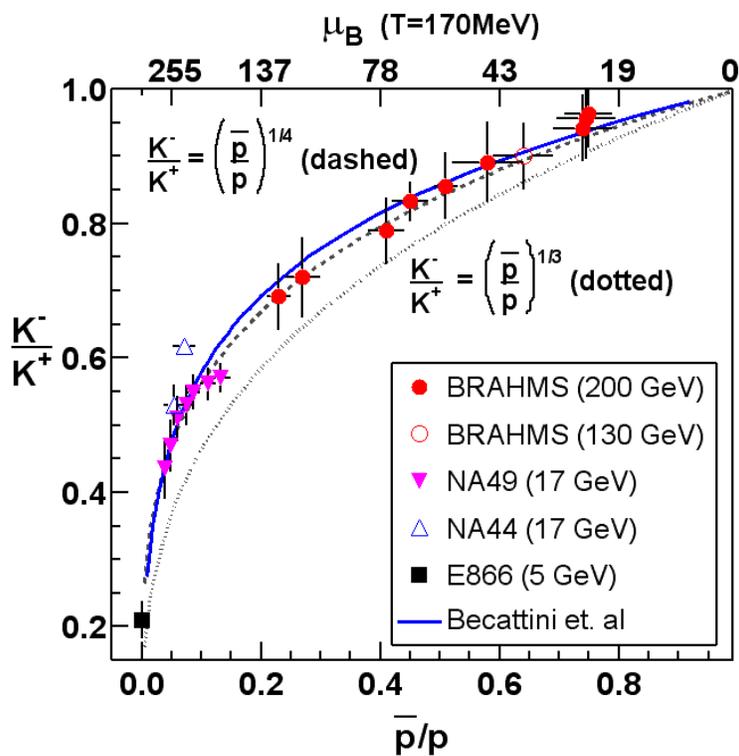

Figure 6. Correlation between the $K^-/K^+$ and antiproton/proton ratios [35].

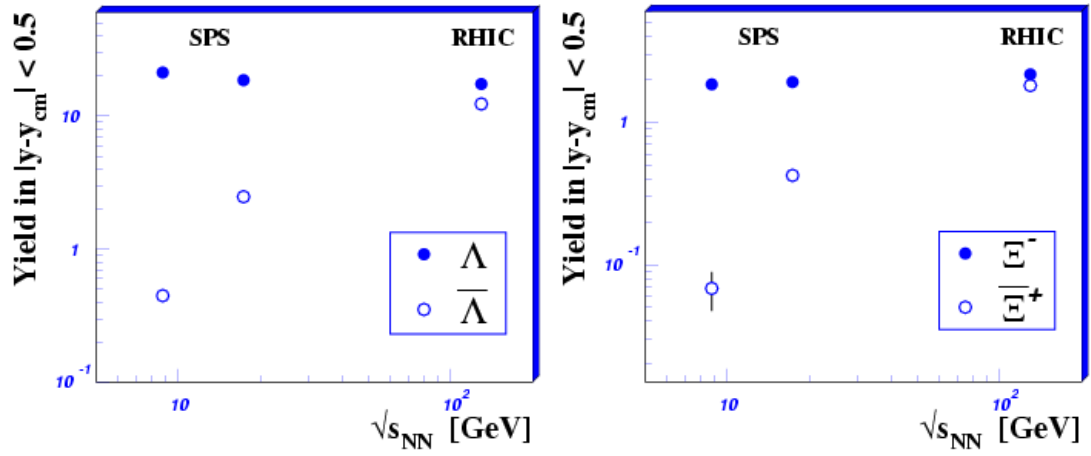

Figure 7. Absolute values of the Λ and Ξ mid-rapidity yields for heavy-ion collisions at SPS (central Pb-Pb) and RHIC (central Au-Au) (see [38] for details).